\documentclass[secnumarabic,amssymb, nobibnotes, aps, prd]{revtex4}%
\usepackage{graphicx}
\usepackage{dcolumn}
\usepackage{bm}
\usepackage{amsmath}%
\usepackage{amsfonts}%
\usepackage{amssymb}
\begin{document}

\title{Muons enhancements at sea level in association with Swift-BAT and MILAGRO triggers}

\author{C. R. A. Augusto, C. E. Navia and K. H. Tsui}
\address{Instituto de F\'{\i}sica, Universidade Federal Fluminense, 24210-346,
Niter\'{o}i, RJ, Brazil} 

\date{\today}
\begin{abstract}
Recently, triggers occurring during high background rate intervals have been reporter by Swift-BAT Gamma Ray Burst (GRB) 
detector. Among them, there were two on January 24, two on January 25, and two on February 13, and 18, all in 2008. 
These Swift-BAT triggers in most cases are probably noise triggers that occurred while 
Swift was entering the South Atlantic Anomaly (SAA). In fact, we show that they happen during a plentiful precipitation
of high energy particles in the SAA, producing muons in the atmosphere detected by two directional telescopes at sea level, inside the SAA region (Tupi experiment). They look like sharp peaks in the muon counting rate. In the same category are two triggers from MILAGRO ground based detector, on January 25 and 31, 2008 respectively. 
In addition, the trigger coordinates are close to (and, in two cases, inside) the field of view of the telescopes.
From an additional analysis in the behavior of the muon counting rate, it is possible to conclude that the events are produced by precipitation of high energy charged particles in the SAA region. Thus, due to its localization, the Tupi experiment constitutes a new  
sensor of high energy particle precipitation in the SAA, and it can be useful in the identification of some triggers of Gamma Ray Burst detectors.
\end{abstract}

\pacs{PACS number: 94.20.wq, 91.25.Rt, 95.85.Ry, 96.50.S-}

\maketitle

\section{Introduction}

The Earth is surrounded by an almost spherical magnetosphere, which shields the Earth against high particle flux. Specially in the day side, the supersonic 
solar wind interaction with the magnetosphere produces a bow shock shielding the Earth of several types of radiations 
(charged particles with up to several GeV energies) coming from external space.  
However, at a certain location over the south Atlantic ocean, centered in the south of Brazil whose equatorial coordinates are 26S and 52W, the shielding effect of the magnetosphere is not quite spherical but shows a hole, as a result of the eccentric displacement of the center of the magnetic field from the geographical center of the Earth
 \cite{frasier87}, this is the so called South Atlantic Anomaly (SAA) region. According to the data of the particle background monitor, aboard of ROSAT spacecraft, the SAA spans from $-50^0$ to $0^0$ geographic latitude and from $-90^0$ to $40^0$ longitude. This behavior of the magnetosphere is responsible for several processes, such as the trapped and azimuthally drifting energetic particles, bouncing between hemispheres, coming deeper down into the atmosphere owing to the low field intensity over SAA, thereby interacting with the dense atmosphere resulting in ionization production and increasing electric conductivity. There is also an enhanced zonal electric field known as the pre-reversal electric field enhancement (PRE) \cite{abdu77,basu01,abdu05}. The PRE arises from combined effects of the eastward thermospheric wind and the longitudinal conductivity gradient. The PRE presents an enhancement at evening hours.
In addition, the open magnetosphere propitiates the magnetic reconnections of
the IMF lines that will take place in this site in the day side \cite{debrito05}. These factors are responsible for an unusually large particle flux, present in the SAA region including particles (protons-ions) with energies above the pion production threshold.
In Fig.1, it is shown the geomagnetic field intensity distribution, represented by iso-intensity lines over the globe.
The lowest value of magnetic intensity situated in southern Brazil defines the position of the SAA region. 

On the other hand, since April 2007 we have been operating the phase II of the Tupi experiment, with two directional muon telescopes at sea level located at 22S and 43W. These coordinates are inside the SAA region and close to its center. This characteristic supplies the muon telescopes the lowest rigidity of response to cosmic protons and ions 
($\geq 0.4$ GV). This value is approximately half of the rigidity at the polar regions.
The telescopes were constructed on the basis of plastic scintillators. One of them has a vertical orientation, and the other is oriented near 45 degrees to the vertical (zenith) pointing to the west, as shown in Fig.2. Both with an effective aperture of $65.6\;cm^2\;sr$. 
The directionality of the vertical muon telescope is guaranteed by a veto or anti-coincidence guard, using a detector of the inclined telescope and vice-verse. Therefore, only muons with trajectories close to the telescope axis are registered. 
The data acquisition is made via software by using a card with an analogical to digital conversion rate of up to 100 kHz. The Tupi experiment has a fully independent 
power supply, with an autonomy of up to 6 hours to guard against eventual power failures. As a result, the data acquisition is carried out during 24 hours, giving a duty cycle higher than 90\%. The telescopes are inside a building under two flagstones of concrete, allowing registration of muons with $E_{\mu}\geq 0.2\;GeV$, required to penetrate the two flagstones (see Fig.2).

The main effect of the SAA is an increase of the muon intensity at sea level, some time in up to 10 times in the day side, 
the precipitation of high energy particles on most of cases begins three hours after the sunrise and finish one hour after the sunset. However this behavior is subject to seasonal variations. We have observed that there is a correlation between the hourly variation of the PRE and the hourly variation of the muon intensity at ground as is shown in Fig.3, in which the upper panel shows the pre-reversal electric field (PRE)
 obtained by simulation under several conductivity gradients \cite{abdu05}, and the lower panel shows the hourly variation of the muon intensity (averaged in two months April and May, 2007). Both panels present the so-called sunset enhancement. It is interesting to observe
that the muon intensity in the vertical telescope is twice the muon intensity in the inclined telescope, in agreement with the expected zenith angle dependence for the muon intensity related as $\cos(\theta)^2=0.5$ for $\theta=45^0$ in the $\sim$GeV region \cite{judge65}. The low rigidity of response of the telescopes is probably responsible for the observation of transient solar events of small scale, already reported earlier \cite{navia05,augusto05}.

The precipitation of high energy particles in the SAA region is subject to fluctuations. For instance, fluctuations exist over several days when no precipitation of particles is observed in this region. The origin of these fluctuations can be described using the ``Open Magnetosphere'' Model \citep{dungey61,debrito05} based on a "reconnection process" 
that takes place at the front (day-side) of the magnetosphere to produce field lines with one end at the Earth and the other in distant space.
The magnetic reconnections of the Interplanetary Magnetic Field (IMF) take place in the SAA region. However, the coupling and specially the decoupling process, under which the IMF evolves and carries out the task of transporting particles out of the SAA region, is still not very well understood. 

On the other hand, Swift is a multi-wavelength space-based observatory dedicated to the study of gamma-ray burst (GRB) science. Its three instruments work together to observe GRBs and their afterglows in the gamma-ray, X-ray, ultraviolet, and optical wavebands. The satellite was developed by an international consortium from the United States, United Kingdom, and Italy \cite{gehrels04}.
The Burst Alert Telescope (BAT) is a 15 - 150 keV energy range detector, and it detects a new GRB event and computes its coordinates in the sky.
MILAGRO experiment is a large-area water-Cherenkov ground-based detector consisting of a 6 million gallon artificial pond instrumented with 723 photomultiplier tubes, and surrounded by an array of 170 individual water-Cherenkov detectors covering an area of approximately 40,000 sq. meters \cite{atkins05,abdo07}. MILAGRO detects gamma-ray and cosmic-ray 
initiated air-showers, and it is sensitive to primary energies of approximately 100 GeV and higher. It is capable of continuously monitoring approximately 2 sr of the overhead sky, allowing us to reconstruct the direction of a gamma-ray initiated extensive air-shower, and search the sky for GRBs. MILAGRO is located at New Mexico (USA) near Los Alamos laboratory, with coordinates 35.9N and 106.7W. 

On January 24, 2008 a new and faster micro-computer has been used for the data acquisition system of our muon telescopes, as well as an up-grade in the data acquisition software. The result is an increment in the counting rate by at least ten-fold. Under these new conditions, we have observed the fine structures of the high energy particle precipitation in the SAA, as hits of cosmic rays producing muons in the Earth atmosphere, observed at ground level by the Tupi telescopes as sharp peaks in the muon counting rate. It is interesting to observe that, in some occasions, the muon peaks, produced by the high energy particle precipitation, coincide with the occurrence of Swift-BAT and MILAGRO triggers. In addition the trigger coordinates are close to, or inside the field of view of the telescopes.
 The Swift-BAT and MILAGRO triggers are obtained from GCN report \cite{barthelmy}, the Swift are: on January 24, 2008 at 15:27:41.15 UT (trigger=301687) and at 15:44:56.54 UT (trigger 301689), on January 25, two similar Swift-BAT triggers at 15:50:48.35 UT 
(trigger=301771) and at 19:11:20 UT (trigger=301779) and so far, on February 13, and 18, 2008 at 14:21:48 UT (trigger=303172) and at 20:08:42 UT
(trigger=303609) respectively. 
In addition, the MILAGRO triggers are: on January 25, 2008 at 18:32:25 (trigger=1167) and on January 31, 2008 at 15:5819 (trigger=28). 

These triggers, at least those observed on January 24, 2008, in the Swift spacecraft according to the CGN CIRCULAR NUMBER 7213 \cite{barthelmy} are considered as noise triggers that occurred while Swift was entering the South Atlantic Anomaly (SAA). 
Meantime, for the other ones, until 
February 22, 2008, the S/C (Swift team) has yet to make its decision about safe to slew 
or not these triggers. Specially, the Swift trigger 303609  due to the Sun constrain, it is almost in coincidence with our sunset.  
In addition, we show that the MILAGRO trigger coordinates point to the SAA region. Details of these results are shown in Fig.4, where the muon counting rate (raw data) in chronological order is shown. Both the Swift-BAT and MILAGRO triggers (indicated as vertical arrows) happen during a plentiful precipitation of high energy particles in the SAA, producing muon enhancements at sea level.

As has already been commented, these Swift-BAT and MILAGRO triggers have their coordinates close to (two of them inside) the field of view of the telescopes. In addition the sharp peaks in the muon counting rate are observed by both telescopes. This means that particle precipitation in SAA  
is constituted by a flow of charged particles, subject to drift processes.
The telescope's axis coordinates, together with the trigger coordinates are plotted in Fig.5 for three cases. The upper panel corresponds to the MILAGRO trigger 1167, the middle panel correspond to the Swift-BAT trigger 303172, and the lower panel corresponds to the Swift-BAT trigger 303609. 
The Swift spacecraft has the Swift-XRT coordinates images, which helps to identify a very bright gamma ray burst from a cosmic ray hit \cite{swift}. While in the MILAGRO case, our results indicate that the MILAGRO detector, even located in the north hemisphere (latitude 35.9N), can be triggered by cosmic ray precipitations in SAA. 
Especially, it can be triggered between 9 UT and 21 UT, when it is pointed to the SAA region, events with negative declination, or even with a positive (but small) declination, due to drift processes.

These results show that the Tupi experiment, due to its localization, constitutes a new  
sensor of high energy particle precipitation in the SAA, and it can be useful in the identification of some triggers of Gamma Ray Burst detectors.  
The Tupi experiment is in the process of expansion, and within the next six months we are able to increase the number of telescopes to fourteen, increasing in 85\% the field of view of the Tupi experiment. 
Our objective is to take advantage of our location inside the anomaly to detect gamma ray burst in the range of GeV to TeV, because there are theoretical models that predict GRBs in this energy region \cite{totani96,dermer00,pilla98}, with strong astrophysical implications. Naturally, we will continue to study solar transient phenomena, and monitoring the precipitation of particles. In the next days, the Tupi data will be in public domain on the Web.

This work is supported by the National Council for Research (CNPq) of Brazil, under Grant  $479813/2004-3$ and $476498/2007-4$. 
We are grateful to various catalogs available on the web and to their open data policy, especially to the GCN report.
For comment and suggestions, please write to e-mail: navia@if.uff.br.

%%%%%%%%%%%%%%%%%%%%%%%

\newpage
%%%%%%%%%%%%%%%%%%%%%%%%%%%%%%%%%%%%%%%%%%%%%%%%%%%%%%%%%%%%%%%%%%%%%%%%%%%%%%%%%%%%%%%% 
\begin{figure}[th]
\vspace*{-0.0cm}
\hspace*{-2.0cm}
\includegraphics[clip,width=1.0
\textwidth,height=1.0\textheight,angle=0.] {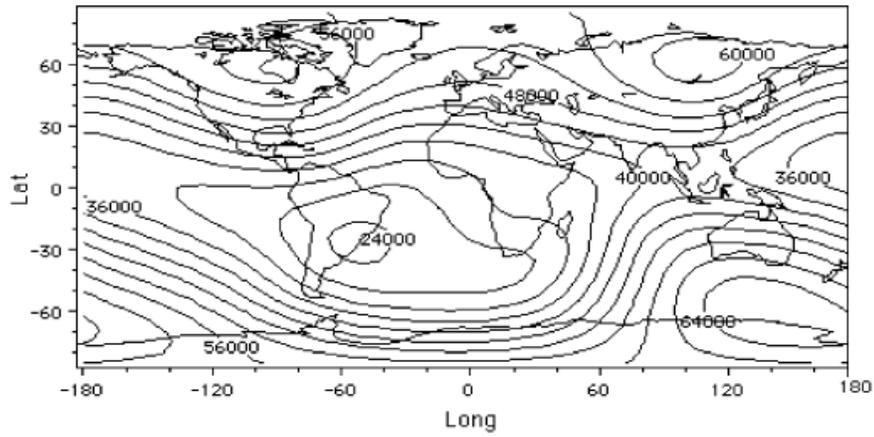}
\vspace*{-15.0cm}
\caption{The geomagnetic total field intensity distribution, represented by iso-intensity lines over the globe.
The lowest value of magnetic intensity situated in South Brazil define the position of the SAA region.}
\end{figure}  
 
\begin{figure}[th]
\vspace*{-1.0cm}
\includegraphics[clip,width=0.6
\textwidth,height=0.6\textheight,angle=0.] {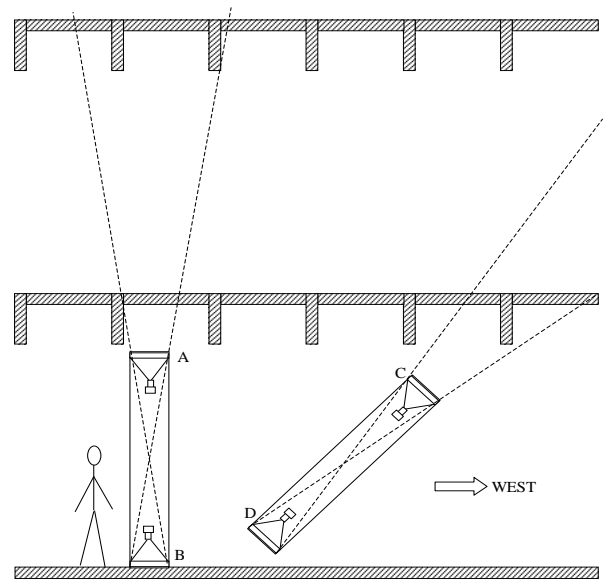}
\vspace*{-4.0cm}
\caption{Experimental setup of the Tupi experiment Phase II, showing the two telescopes.}
\end{figure} 

\begin{figure}[th]
\vspace*{+1.0cm}
\hspace*{-3.0cm}
\includegraphics[clip,width=1.0
\textwidth,height=1.0\textheight,angle=0.] {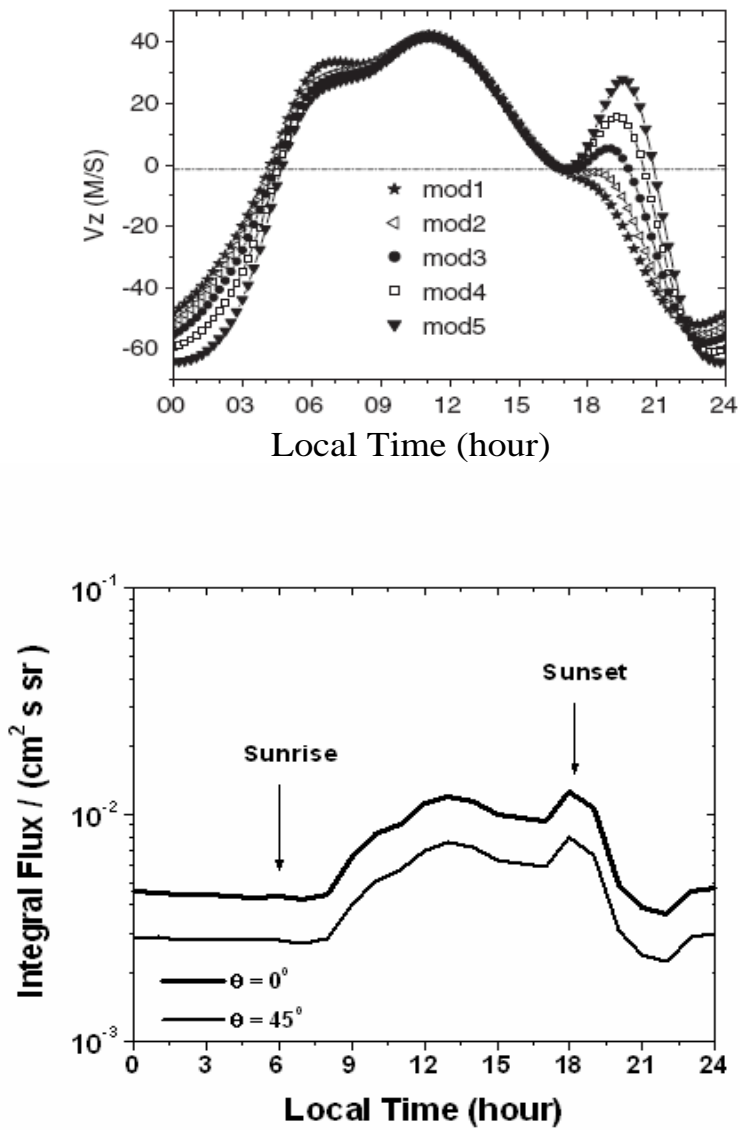}
\vspace*{-7.0cm}
\caption{Upper panel, Monte Carlo results of the hourly variation of the atmospheric pre-reversal electric field (PRE). The curves identified as mod 1 and mod 5 correspond to the lowest and highest conductivity gradients at sunset \citep{abdu05}. Lower panel, (two month averages April-May, 2007) hourly variation of the muon intensity  observed by Tupi telescopes.}
\end{figure}

\begin{figure}[th]
\vspace*{+1.0cm}
\hspace*{-1.0cm}
\includegraphics[clip,width=0.8
\textwidth,height=0.8\textheight,angle=0.] {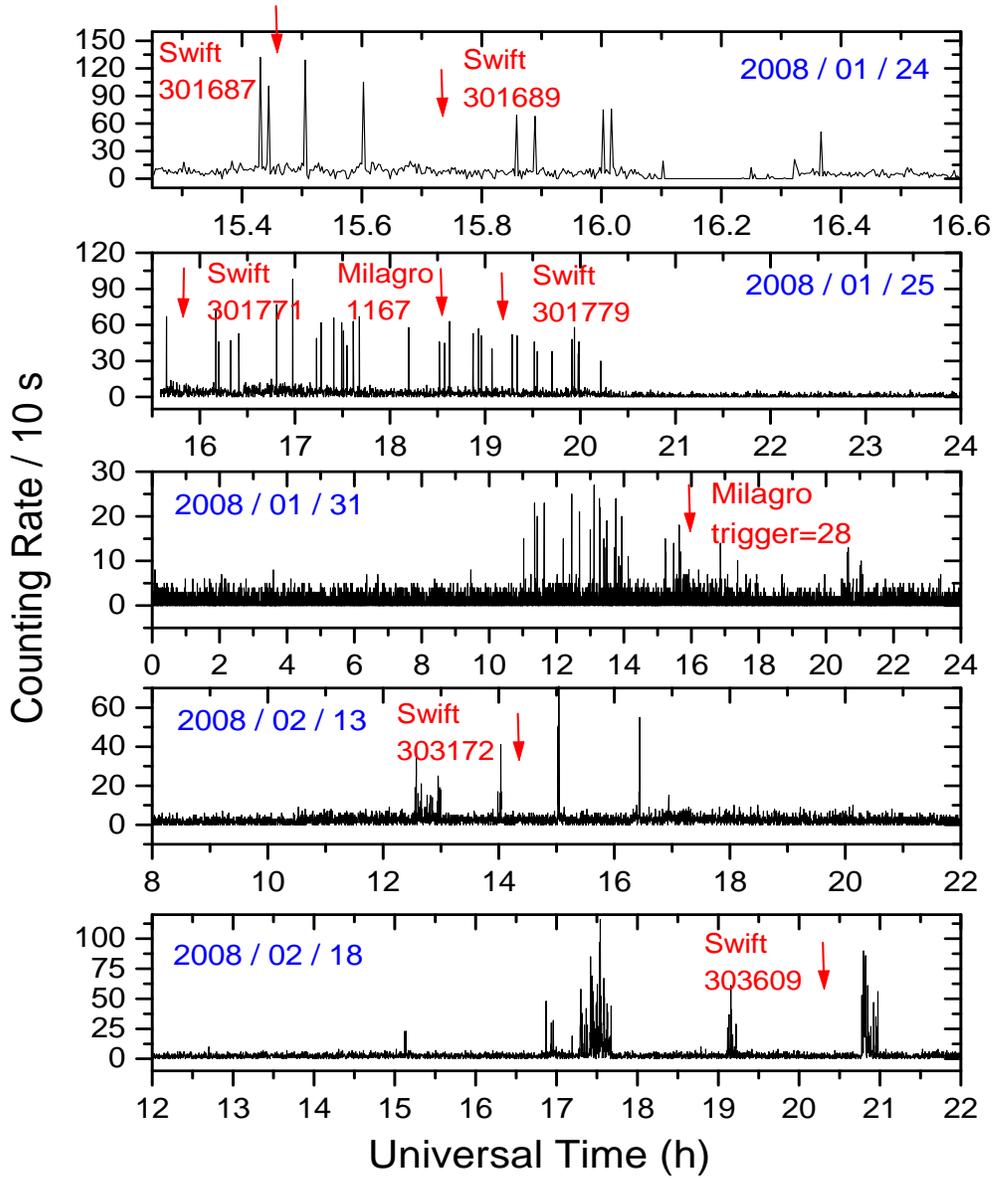}
\vspace*{-1.0cm}
\caption{Precipitation of high energy particles in the SAA region and observed as sharp peaks in the muon counting rate of the vertical Tupi telescopes on January 24, 25, and 31, 2008, and Febreuary 13, and 18, 2008, respectively. The vertical arrows represent the Swift-BAT and MILAGRO trigger occurrences.}
\end{figure}

\newpage

\begin{figure}[th]
\vspace*{+1.0cm}
\hspace*{-0.0cm}
\includegraphics[clip,width=0.6
\textwidth,height=0.6\textheight,angle=0.] {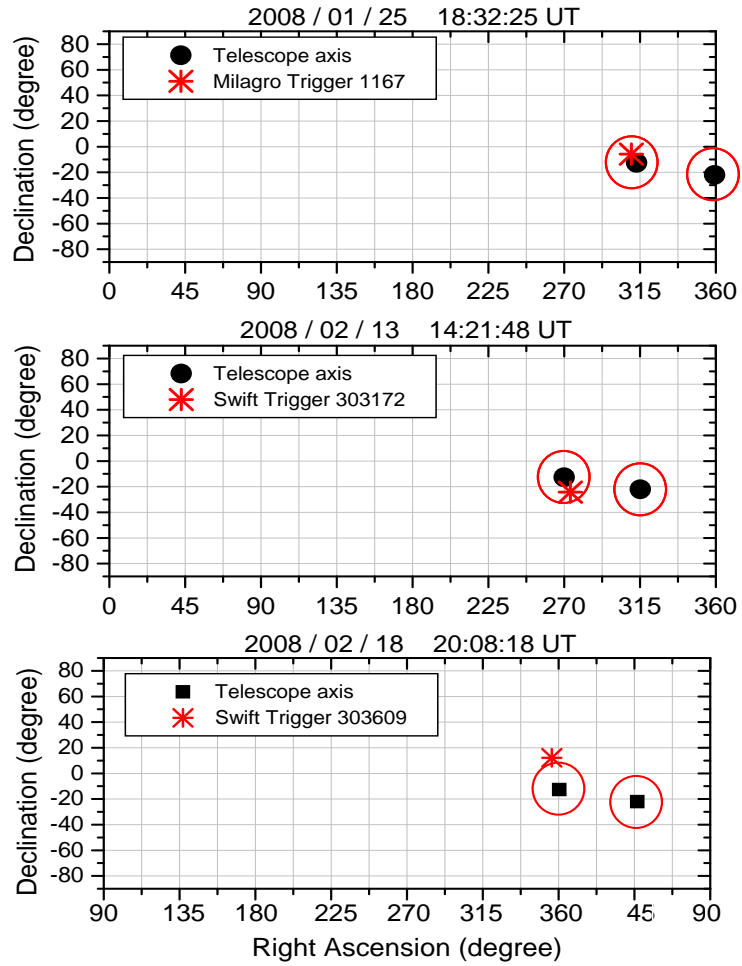}
\vspace*{-0.0cm}
\caption{Equatorial coordinates showing the position of the two telescopes axis, the ``circles'' represent the field of view of the telescopes and the the asterisk is the position (coordinates) of the: Upper panel the MILAGRO trigger 1167, middle panel the Swift-BAT trigger 303172 and the 
lower panel the Swift-BAT trigger 303609.}
\end{figure}

\end{document}